\documentclass[fleqn,usenatbib]{mnras}

\usepackage{newtxtext,newtxmath}
\usepackage{verbatim}

\usepackage[T1]{fontenc}
\usepackage{ae,aecompl}
\usepackage[svgnames]{xcolor}
\usepackage{hyperref}
\hypersetup{citecolor=Blue}

\usepackage{graphicx}	
\usepackage{amsmath}	
\usepackage{amssymb}	
\usepackage{amsfonts}
\usepackage{epsfig,rotating}

\def\simeq{\mathrel{\raise.3ex\hbox{$\sim$}\mkern-14mu\lower0.4ex\hbox{$-$}}}

\def\ltsima{$\; \buildrel < \over \sim \;$}
\def\simlt{\lower.5ex\hbox{\ltsima}}
\def\gtsima{$\; \buildrel > \over \sim \;$}
\def\simgt{\lower.5ex\hbox{\gtsima}}

\def\msun{{\rm M_{\odot}}}

\def\be{\begin{equation}}
\def\ee{\end{equation}}

\def\del#1{{}}

\newcommand\mearth{\,{\rm M}_{\oplus}}
\newcommand\mj{\,{\rm M}_{\rm J}}

\title[Slow runaway accretion]{ALMA observations require slower Core Accretion runaway growth}

\author[Nayakshin, Dipierro \& Szul\'agyi]{S. Nayakshin$^{1}$, G. Dipierro$^{1}$ and J. Szul\'agyi$^{2}$\\
$^{1}$ Department of Physics and Astronomy, University of Leicester, University Road, LE1 7RH Leicester, United Kingdom\\
$^{2}$ Center for Theoretical Astrophysics and Cosmology, Institute for Computational Science, University of Z\"urich, Winterthurerstrasse 190, CH-8057 Z\"urich,\\ Switzerland
}

\date{Accepted XXX. Received YYY; in original form ZZZ}

\pubyear{2019}
\hypersetup{draft}

\begin{document}
\label{firstpage}
\pagerange{\pageref{firstpage}--\pageref{lastpage}}
\maketitle

\begin{abstract}
Thanks to recent high resolution ALMA observations, there is an accumulating evidence for presence of giant planets with masses from $\sim 0.01 \mj$ to a few $\mj$ with separations up to $ 100$~AU in the annular structures observed in young protoplanetary discs. We point out that these observations set unique "live" constraints on the process of gas accretion onto sub-Jovian planets that were not previously available.  Accordingly, we use a population synthesis approach in a new way: we build time-resolved models and compare the properties of the synthetic planets with the ALMA data at the same age. Applying the widely used gas accretion formulae leads to a deficit of sub-Jovian planets and an over-abundance of a few Jupiter mass planets compared to observations. We find that gas accretion rate onto planets needs to be suppressed by about an order of magnitude to match the observed planet mass function. This slower gas giant growth predicts that the planet mass should correlate positively with the age of the protoplanetary disc, albeit with a large scatter. This effect is not clearly present in the ALMA data but may be confirmed in the near future with more observations. 
\end{abstract}




\section{Introduction}\label{sec:intro}

In the Core Accretion paradigm, a solid core grows by accretion of solids \citep{Safronov72,PollackEtal96}.  When the core mass reaches a critical value of $M_{\rm cr}\sim 10-20\mearth$, a gas envelope collapses onto the core, fuelling a phase of rapid gas accretion \citep{MizunoEtal78,IkomaEtal00}, during which the planet accretion rate is believed to be limited only by the rate at which the disc supplies it with gas. Detailed isothermal simulations showed that the planet may grow from the mass of $\sim 0.1\mj$ to $\sim (1-3) \mj$ in a matter of less than ten thousand years \citep{BateEtal03,DangeloEtal03}. In contrast, dispersal of protoplanetary discs takes $\sim 3$ Myr \citep{HaischEtal01}. A planet is hence destined to become a massive gas giant if it enters the runaway accretion growth phase while the disc is still present. 

This  runaway gas accretion scenario produces {\it a valley in the planet mass function} from $M_{\rm p} \sim 0.1 \mj$ to $M_{\rm p} \sim 1 \mj$ or more \citep{IdaLin04a,MordasiniEtal09b}. Early exoplanet observations seemed to confirm this \citep[e.g., see the red histogram in fig. 12 in][]{MayorEtal11}. However, a number of new observational and theoretical arguments suggest that gas accretion onto planets is significantly less efficient than hitherto believed. Non-isothermal multi-dimensional simulations suggest that atmospheric circulation and inefficient radiative cooling may reduce accretion rates significantly \citep{AyliffeBate09,AyliffeBate09b,OrmelEtal15,SzulagyiEtal16,SzulagyiMordasini17,Szulagyi17,CimermanEtal17,LambrechtsLega17}. This may explain why super-Earth planets with very modest gaseous atmospheres are abundant at separations of $\sim 0.1$~AU \citep{LeeChiangOrmel14}.

Additionally, the classical positive correlation of gas giants with host star metallicity \citep{FischerValenti05}, dominated by $\sim 1\mj$ planets, was found to disappear at higher masses \citep{SantosEtal17}. The exact mass scale at which the switch in the metallicity correlations takes place is currently debated but is somewhere between $2-10\mj$  \citep{Schlaufman18,Adibekyan19,MaldonadoEtal19,GodaMatsuo19}. The trend is continuous into the brown dwarf regime \citep[e.g.,][]{TroupEtal16}, and suggests that at least a fraction of the most massive planets forms "as stars" -- by disc fragmentation. This also suggests that gas accretion onto $\gtrsim 1\mj$ mass planets is inefficient as otherwise the positive host metallicity correlation of low mass ``seed'' giants would be passed on to higher mass planets, and even strengthened  \citep[][]{MordasiniEtal12}.

Additional support for these ideas comes from the microlensing surveys sensitive to planets with separations of a few AU \citep{SuzukiEtal16}.  The mass function of microlensing planets  contains too many  planets in the mass range $\sim[0.1 - 0.3] \mj$ compared with the runaway accretion scenario \citep{SuzukiEtal18}. The models were shown to fare better if gas accretion rate onto sub-Jupiter mass planets that open deep gaps in the disc and isolate themselves from the gas supply is reduced. In this scenario, $M_{\rm p} \gtrsim 1\mj$ may preferentially form by  gravitational instability \citep{SuzukiEtal18}.

ALMA observations  of annular structures (rings and gaps) in the dust emission of young protoplanetary discs at $\sim 10-100$~AU separations have been interpreted as signs of gas giant planets by many authors \citep{BroganEtal15,DipierroEtal15,DongEtal18-alpha0,LongEtal18,Dsharp1}.  While other interpretations exist (see Discussion), it is important to ask what these candidate planets may mean for planet formation theories. \cite{LodatoEtal19} (L19 hereafter) considered evolutionary paths of these candidate planets, starting from their observed parameters,  and found that they evolve into massive gas giants quickly. We ask a complementary question: {\em how did the candidate planets evolve to be what they are now}? These $\sim 1-10$ Million years old gaseous discs have not yet been dispersed and should fuel planetary growth, giving us the first ever time-resolved observational probe of runaway gas accretion. Furthermore, at large separations the runaway accretion should start at lower core masses  \citep{Piso15} and terminate at larger masses ($\sim [3 - 10]\mj$, because disc gap-opening is harder), producing a very wide  valley from $\sim 0.1 \mj$ to $\sim 3\mj$, and an excess of $M_{\rm p}\sim 10\mj$ planets. 


\begin{figure*}
\includegraphics[width=0.99\textwidth]{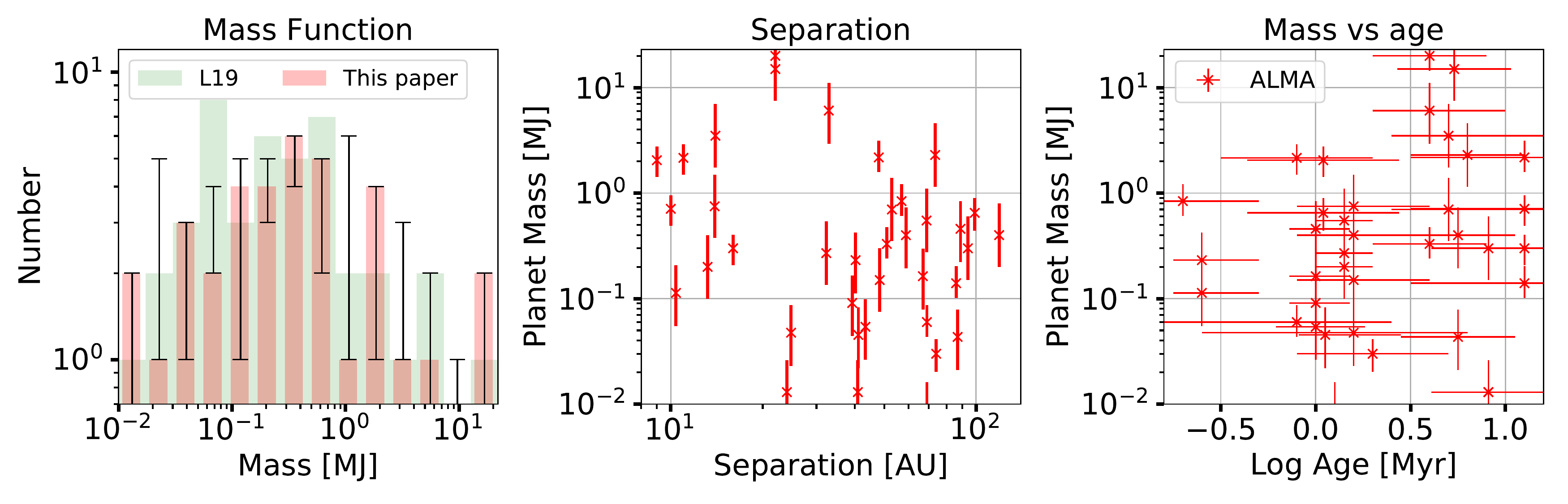}
\caption{Data from Table \ref{table1}. {\bf Left:} The candidate planet mass histogram from Table \ref{table1}, shown with the red color and the error bars. The green histogram shows the planet masses from L19 for comparison; {\bf Middle:} Mass versus separation; {\bf Right:} Planet mass versus age of the system.}
\label{fig:DATA}
\end{figure*}

\section{The data}\label{sec:data}

Table \ref{table1} shows the masses, separation and ages of the ALMA candidate planets that we use here. Our two major data sources are \cite{LongEtal18} and the DSHARP \citep{Dsharp1,Dsharp2} survey. The former lists dust gap widths $W$, separations, $a$, stellar masses, $M_*$, and other relevant information. Following \cite{LodatoEtal19},  we assume that the gap width scales with the planet Hill radius.
Most of the hydrodynamical simulations {\bf to date}  have shown that $W$ corresponds to the range $ (4.5 - 7) R_{\rm H}$, where $R_{\rm H} = a (M_{\rm p}/3 M_*)^{1/3}$ is the Hill radius for the planet of mass $M_{\rm p}$  \citep[e.g.,][]{LiuDEtal19}. We assume a reference value of $W = 5.5 R_{\rm H}$ to compute $M_{\rm p}$. The planet mass error bars are determined by solving for the maximum and minimum values of $M_{\rm p}$ via the gap width parameters $w_{\rm max} = 4.5$ and  $w_{\rm min} = 7$, respectively. For several of the objects from the \cite{LongEtal18} sample, e.g., HL TAU and CI TAU, we instead rely on the results of published dedicated hydrodynamical simulations \citep[][]{DipierroEtal15,ClarkeEtal18}. \citet{Dsharp7} lists DSHARP planet masses derived from detailed hydrodynamcal simulations, including error bars. From their Table 3 we use only model $M_{\rm p, am3}$ and DSD1 dust model, and omit the lower part of their table since most of the estimated masses there are consistent with 0 within the errors. For additional sources from other publications, TW Hya \citep{MentiplayEtal18}, GY 91 \citep{SheehanEisner18}, HD 169142 \citep{PohlEtal17}, PDS 70 \citep{KepplerEtal19}, we assume that the planet mass error is $\pm\log_{10} (2)$.


Fig. \ref{fig:DATA} shows the resulting planet mass histogram, the candidate planet mass versus separation, and mass versus age of the system. Our mass histogram  (pale red with error bars) is rather similar to that derived by L19 (shown with the green colour).

\section{Population synthesis modeling}\label{sec:no_migration}

In brief, to compare theoretical predictions to observations, we accept a simplified model for the protoplanetary disc structure and evolution. A massive core is then injected into the disc and allowed to grow and undergo the runaway gas accretion growth at rate $C^{-1} \dot M_{\rm ra}$, where $C \ge 1$ is a dimensionless factor, and $\dot M_{\rm ra}$ is the runaway rate given by one of three runaway models from the literature described below. The population synthesis is terminated at the age of the ALMA systems and the resulting planet masses are compared to those of the candidate ALMA planets. A reasonable agreement is found for $C\sim 10$.

We model the disc surface density as a power law in radius with a time dependent normalization, 
\begin{equation}
    \Sigma(R, t)  = \frac{M_{\rm d}(t)}{2\pi R (R_{\rm out} - R_{\rm in})} \;,
\end{equation}
where $M_{\rm d}$ is the disc mass between  radii $R_{\rm in} = 5$~AU and $R_{\rm out} =100$~AU. 
The disc mass is evolved according to
\begin{equation}
\frac{d M_{\rm d}}{dt} = -\frac{M_{\rm d}}{t_{\rm disp}} - \dot M_{\rm g}\;,
\end{equation}
where $\dot M_{\rm g}$ is the gas accretion rate onto the planet, and the disc dispersal time is $t_{\rm disp} = 2$ Myr. The initial disk mass is a uniform random variable with the minimum and maximum disc masses of $0.025\msun$ and $0.1\msun$, respectively, whereas stellar mass is $M_* = 1\msun$. The disc is in a vertical hydrostatic equilibrium, i.e. the geometric aspect ratio is $H/R = \sqrt{k_b T R/(GM_* \mu)}$, where $k_b$ is the Boltzmann's constant, $G$ is the gravitational constant, $\mu = 2.45 m_p$ is the mean molecular weight where $m_p$ is the proton mass, and the disc midplane temperature $T$ is given by $T = T_0 (R_0/R)^{1/2}$ where $T_0 = 20$~K and $R_0 = 100$~AU. We fix the disc $\alpha$-viscosity parameter \citep{Shakura73} at $\alpha=0.005$. The value of $\alpha$ affects the disc gap opening condition, for which we follow \cite{CridaEtal06} results, and may also affect gas accretion rates onto the planets depending on the exact scenario described below.


A growing core is injected into the disc at separation $R$, where $R$ is a random variable with a uniform distribution in the $\log R$ space between $R=10$~AU and $R=140$~AU. We neglect planet migration for clarity of the argument. The classical runaway accretion phase duration is as short as $\sim 10^4$ years \citep{DangeloEtal03}, implying that the orbital separation of the planet will shrink by a small fraction only during the runaway phase. For example, \cite{DipierroEtal18} model the rings and gaps in the dusty disc of Elias 24 (one of the systems included in our paper). Their planet grows from mass of $0.15 \mj$ to $0.7 \mj$ in about $4.4\times 10^4$ years, during which it migrates from 65 AU to 61.7 AU only.  Population synthesis with the three widely known models also shows that planets born beyond tens of AU do not start to migrate appreciably  until the mass of $1-3\mj$ \citep[e.g., see Fig. 3 in][]{IdaEtal18}, which is more massive than most of the ALMA planets.

We tested two methods of injecting massive solid cores into the disc and obtained very similar results. In the first, the cores are injected into the disc with mass large enough (e.g., $10-20 \mearth$) for the runaway accretion to start immediately, and the core injection time,  $t_0$, was a random variable distributed uniformly in the log space between $0.1$~Myr and $1$~Myr. In the second method, presented in the paper, the cores are injected in the disc at time $t_0 = 0$ with smaller initial mass, $M_{\rm c} = 1 \mearth$. The cores then accrete solids at a rate, $\dot M_{\rm c}$, that is a uniform random variable in the $\log$ space in the limits between $3\times 10^{-6} \mearth$~yr$^{-1}$ and $3\times 10^{-5} \mearth$~yr$^{-1}$. We found that choosing larger values of $\dot M_{\rm c}$ does not affect our conclusions on $C$ but reduces the number of sub-critical cores which have not yet entered the runaway valley. Choosing $\dot M_{\rm c} \ll 3 \times 10^{-6} \mearth$~yr$^{-1}$ leads to too few massive cores $M_{\rm c} \simgt 10\mearth$, and hence too few gas giants, failing to explain the data. 


The gas envelope of the planet grows at the rate
\begin{equation}
    \dot M_{\rm g} = \frac{M_{\rm p}}{t_{\rm kh}}\;,
\end{equation}
where $t_{\rm kh}$ is the Kelvin-Helmholz timescale of the envelope \citep{IkomaEtal00,IdaLin04a}, which we write as
\begin{equation}
    t_{\rm kh} = 10^3 \hbox{ yr } \left(\frac{100 \mearth}{M_{\rm p}}\right)^{3} \kappa_0\;,
\end{equation}
where $\kappa_0$ is gas dust opacity in units of 1 cm$^2$~g$^{-1}$ \citep[see eq. 10 in][we set $k_0 =1$ in this paper except for the Bern model]{IdaEtal18}. The total accretion rate of solids and gas is the sum $\dot M_{\rm p} = \dot M_{\rm c} + \dot M_{\rm g}$. We terminate accretion of solids when $M_{\rm p} \ge 30\mearth$\footnote{Gas accretion dominates strongly at these masses anyway, but also ALMA observations show that the total disc dust masses are $\sim 30-100 \mearth$ \citep{LongEtal18,DSHARP-6}.}. The above model for planetary growth is capped by the disc-limited runaway gas accretion rate $\dot M_{\rm ra}$ that is different for the three population synthesis scenarios explored below:
\begin{itemize}
    \item In the IL04 model \citep{IdaLin04a,IdaLin04b}, the runaway accretion rate is given by $\dot M_{\rm ra} = \dot M_{\rm d} \exp(-M_{\rm p}/M_{\rm th})$, where $\dot M_{\rm d} = 3\pi \alpha c_{\rm s} H \Sigma$ is the unperturbed viscous disc accretion rate, and the thermal mass $M_{\rm th} =120\mearth (R/\hbox{AU})^{3/4}$ takes into account gap opening that assumes to terminate gas supply to the planet.

\item In the Bern model  \citep{MordasiniEtal12a}, the gas runaway accretion rate depends on whether the planet opened a deep gap in the disc or not. In the former case, $\dot M_{\rm d} = 3\pi \alpha c_{\rm s} H \Sigma$. In the latter case, $\dot M_{\rm ra} = \Sigma \Omega R_{\rm gc}^3 H^{-1}$, where $R_{\rm gc}$ is the gas capture radius (their eq. 14). We employ the \citet{CridaEtal06} gap opening condition to select the appropriate limit. Following the authors, we use a reduced dust opacity $\kappa_0 = 0.003$.

\item In the \cite{TT16} (TT16 hereafter), the  runaway gas accretion rate is the minimum of two expressions. The first, gas accretion in the embedded phase, is given by $\dot M_{\rm emb} = 0.29 (M_{\rm p}/M_*)^{2/3} R^4 \Omega \Sigma_{\rm pert} H^{-2}$, where $\Sigma_{\rm pert} = \Sigma(R,t)/(1 + 0.034 K)$ is the disc surface density perturbed by the presence of the planet, with the factor given by $K = (R/H)^5 (M_{\rm p}/M_*)^2 \alpha^{-1}$. The second is the maximum of the global viscous disc accretion rate, $\dot M_{\rm d}$, specified previously, and the "local" rate \citep[eq. 13 in][]{TT16} that is applicable only while the planet is clearing its local gas reserves. We neglect the latter phase for simplicity and note that its inclusion would require an even larger accretion rate suppression.
\end{itemize}
Finally, we apply the suppression factor $C\ge 1$ for all the models and compare the three models with observation. We do not allow the planets to exceed the mass of $10 \mj$ in order to remain in the planetary regime.

\section{Comparison of results to observations}

L19 used population synthesis and asked how the ALMA planet candidates will evolve by the time their discs are dispersed. Here we ask a complementary question: can we form the observed population of ALMA planets within the widely accepted framework for planetary growth? Therefore,  we terminate our population synthesis at the age of the ALMA discs. In practice, the termination time of our models is randomly selected from the list of the stellar ages (shown in Table \ref{table1}), $t_{\rm A}$, multiplied by a random number between 0.5 and 1.5. This multiplication bears no practical importance but improves visibility of population synthesis planet ages in the figures. 

\begin{figure*}
\includegraphics[width=0.33\textwidth]{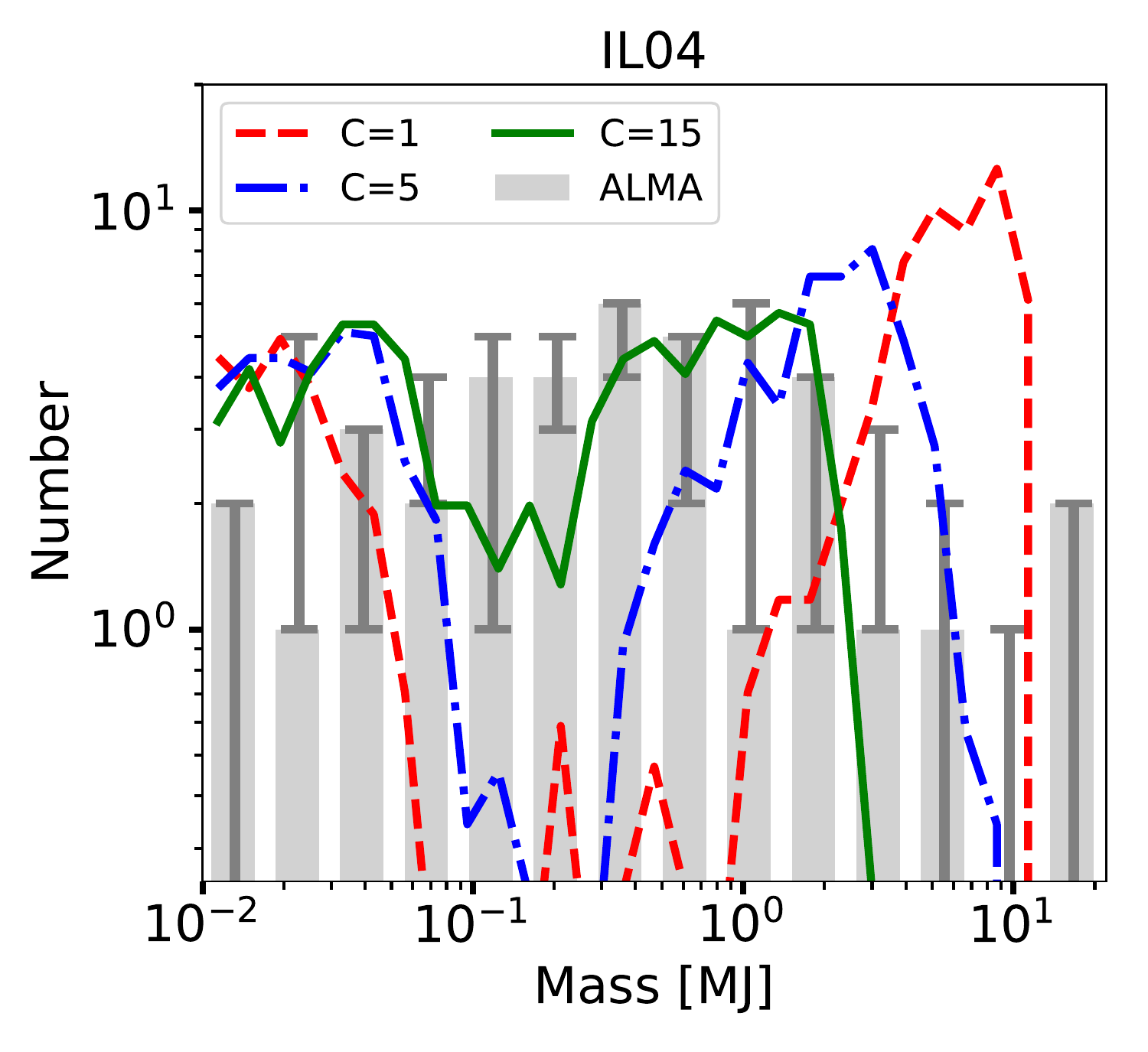}
\includegraphics[width=0.33\textwidth]{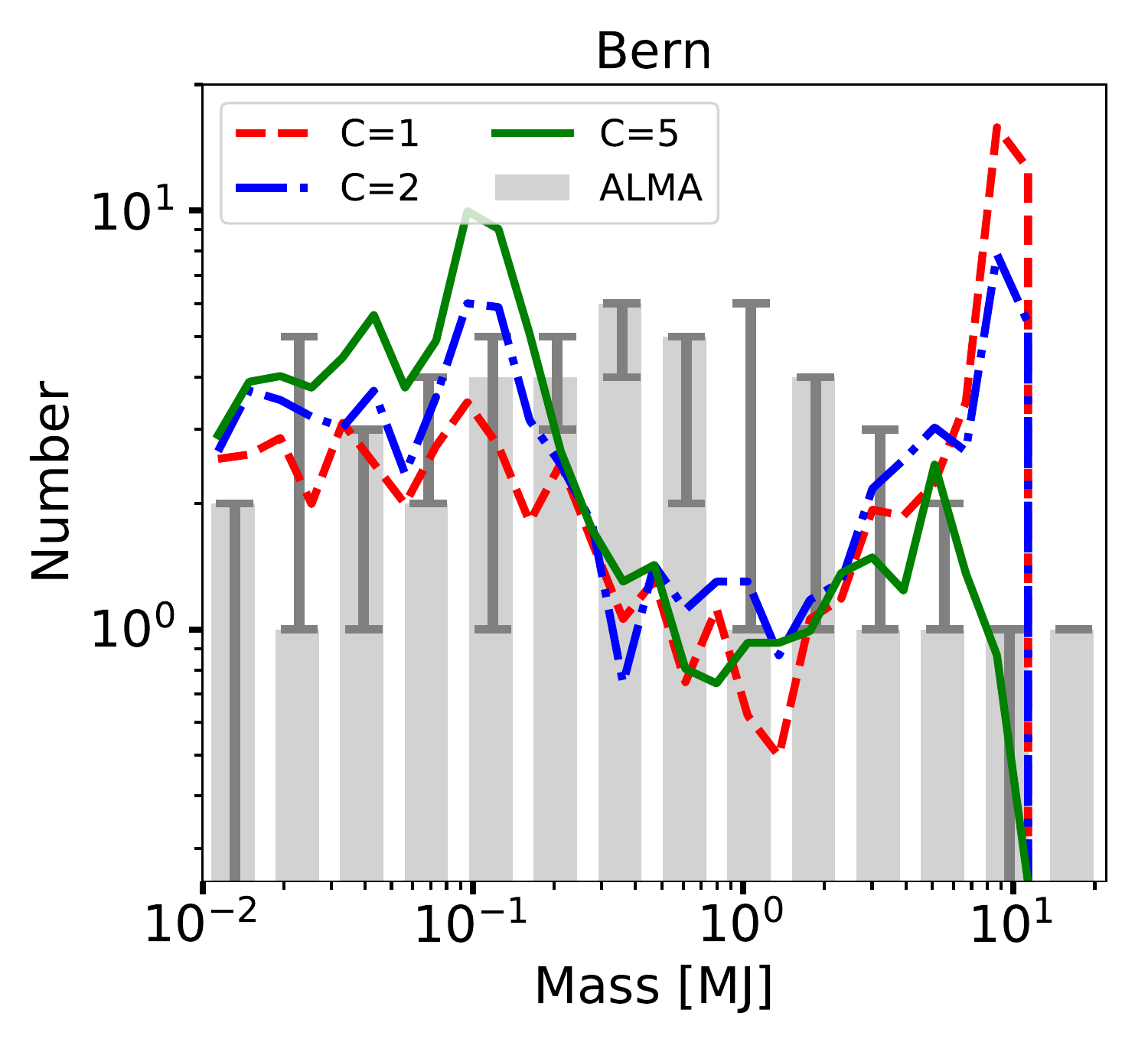}
\includegraphics[width=0.33\textwidth]{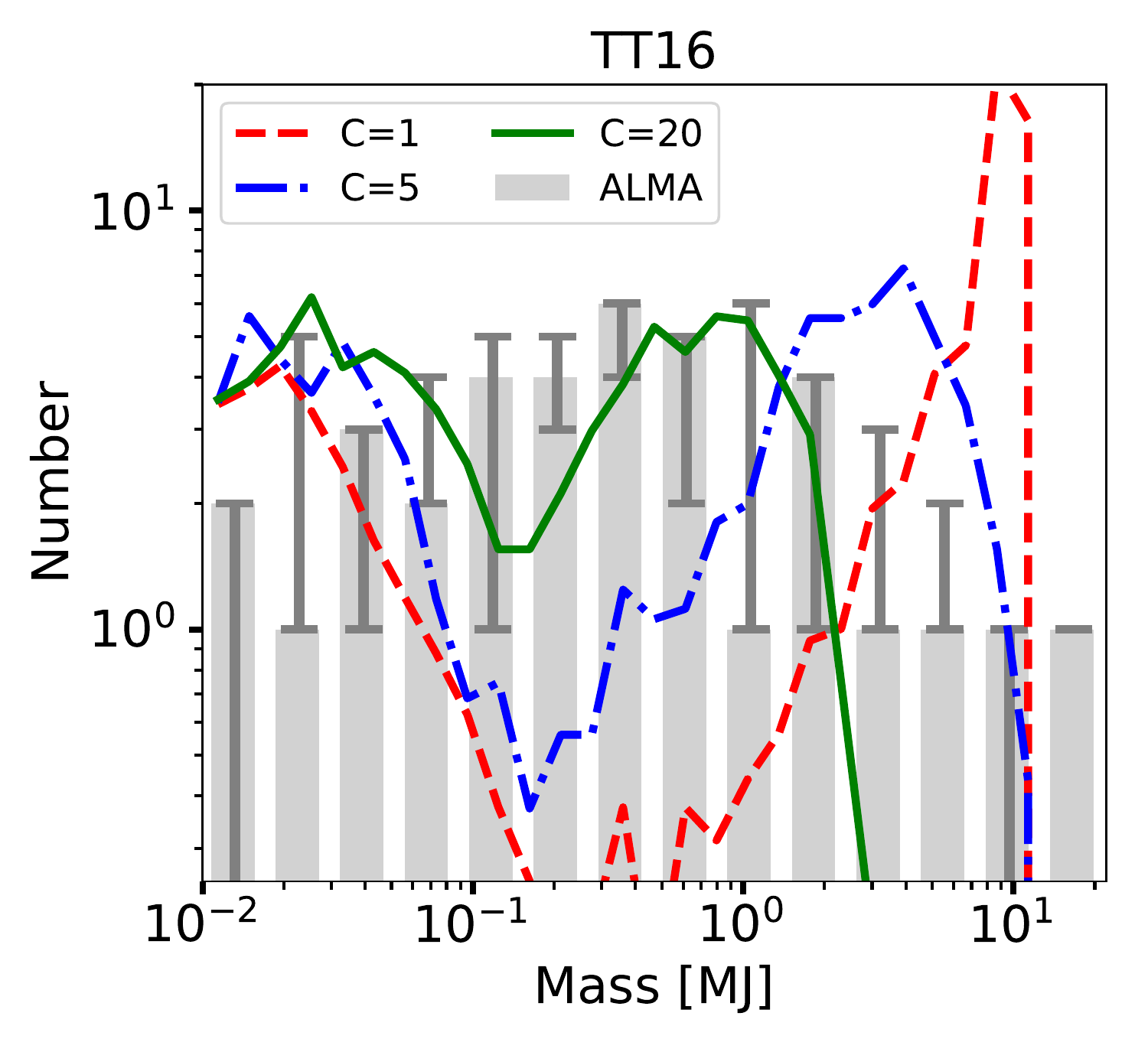}
\caption{The mass function for the three population synthesis models (colored curves) of runaway accretion growth compared to ALMA observations (gray). The normalization of synthetic planet populations was scaled down to match the total number of ALMA planets. Gray capped vertical lines shows estimated observational errors. Note that all of the models over-predict the population of massive gas giant planets strongly for $C=1$. The gas runaway accretion rate needs to be suppressed by the factor of $\sim 5-25$ to yield a reasonable match to the observed mass function.}
\label{fig:MF}
\end{figure*}

Fig. \ref{fig:MF} compares the mass function of ALMA planets, shown with the gray histogram, with the resulting mass distribution of the three population synthesis calculations for various values of $C$. The nominal runaway gas accretion models (red, $C=1$) are inconsistent with the data. The Bern model shows the smallest disagreement since its nominal accretion rate $\dot M_{\rm ra}$ is lower than the two other models. In all cases there is a synthetic planet desert between the mass of $\sim 0.1\mj$ to several $\mj$. The population of very massive gas giants seen in the synthetic models should be easily observable in the ALMA data as such planets open gaps not only in dust but also in gas, but such planets are rare. Suppression of gas accretion rates by factors of $C>1$ reduces the disagreement between the models and the data. Qualitatively, the models with $C=15$, 5 and 20,  for the IL04, Bern and TT16 models, respectively, produce reasonable planet mass functions (solid green curves).

\begin{figure*}
\includegraphics[width=0.45\textwidth]{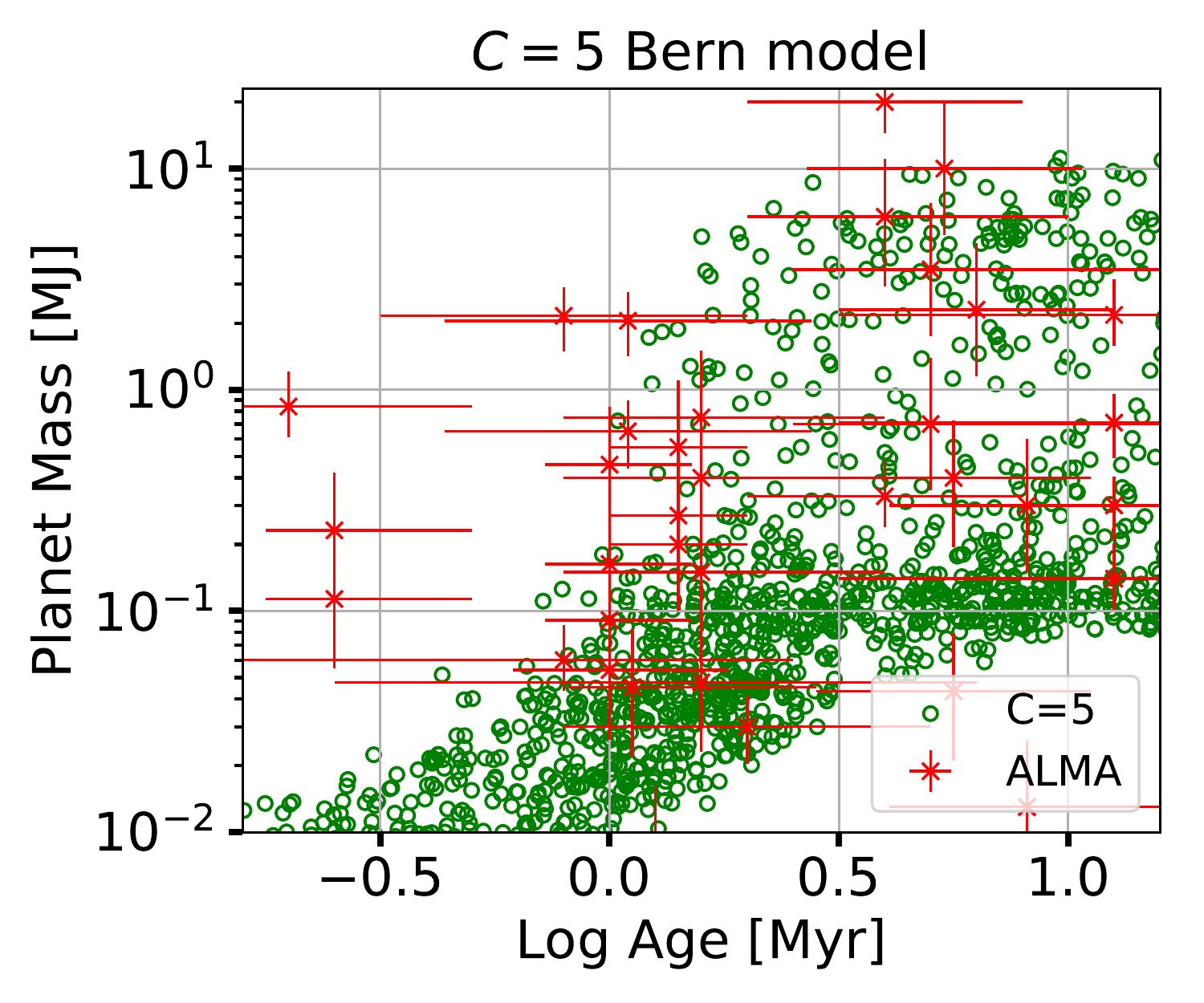}
\includegraphics[width=0.45\textwidth]{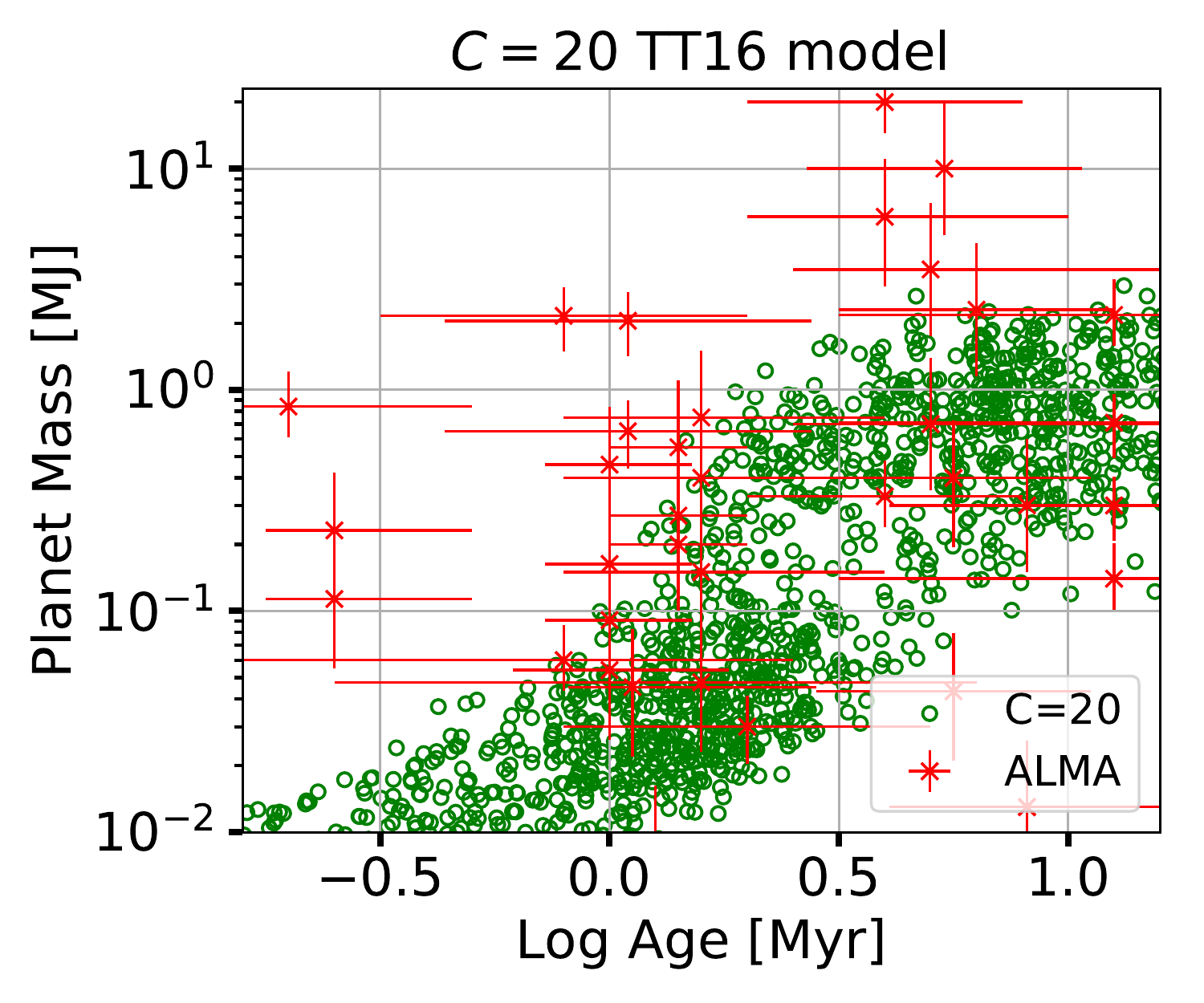}
\caption{Distribution of the ALMA data in the planet mass versus stellar age compared to the two synthetic models with the reasonable values of $C$ as indicated in the legends. This mass-age correlation that may be useful to test theoretical models against the data in the future.}
\label{fig:pop-syn2}
\end{figure*}

Fig.~\ref{fig:pop-syn2} shows the synthetic planet masses versus system ages for two of the models ($C=5$ Bern and $C=20$ TT16 models), compared to ALMA data (red symbols with the  error bars). The synthetic planets show a trend of increasing planet mass with the system age, which is to be expected. Synthetic planets more massive than $1\mj$ are usually older than $\sim 3$ Myrs.  It is not clear if the current ALMA data support or challenge theoretical predictions. There are several massive planets too young for their masses for the population models to reproduce, however the age estimates have significant uncertainties. 


\section{Discussion and Conclusions}

We showed that the runaway gas accretion scenario predicts too few sub-Jupiter mass giants and too many $\gtrsim 3\mj$ planets compared with the candidate ALMA planets. Reduction in the efficiency of gas accretion by a factor of order 10 results in a better agreement of theory and observations. 

Although icelines, dead zone transition, secular gravitational instability and other effects were proposed to explain the observed ringed dust structures {\it instead of} planets \citep{ZhangEtal15-cond-front,PinillaEtal16-Dead-zone,FlockEtal15-MRI,TI14-dust-rings-GI}, these are not likely to account for most of the observations because there is no correlation between the structures and thermal disc properties, and because some of the rings are narrower than the disc scaleheight $H$ \citep{LongEtal18,Dsharp2,Dsharp7}.

Some authors suggested that candidate planets are less massive than the typically inferred $\sim$ Saturn masses, e.g., $\sim 10-20 \mearth$ in \cite{Boley17}; $M_{\rm p} \sim 1-60 \mearth$ in \cite{DongEtal18-alpha0}. However, planets less massive than $\sim 10\mearth$ are unlikely to work for most sources as the disc viscosity parameter required in these scenarios, $\alpha\lesssim 10^{-4}$, is too small to account for the observed ring profiles \citep[see \S 5.3 in][]{DSHARP-6}, and also cannot explain the observed stellar accretion rates, assuming a viscous angular momentum transport in the disc \citep[see][]{ClarkeEtal18}. Furthermore, runaway gas accretion sets in at lower core masses -- as low as $\sim 5\mearth$ at wide separations \citep{PisoYoudin14}. Therefore, the exact planet masses are not crucial for validity of our argument as long as the planets are more massive than $\sim 10\mearth$; the runaway gas accretion would still happen and the results would then diverge from the ALMA observations significantly. There are also suggestions that planets may be larger, e.g., $M_{\rm p} \sim 1\mj$ \citep{BaeEtal18-jupiters}. However, $M_{\rm p}\sim 1 \mj$ planets are inside the ``forbidden'' middle region of the runaway valley. The mere (wide-spread) existence of such planets at large separations would again challenge an unabated runaway growth scenario.  On the other hand, the number of planets in ALMA discs may be smaller since planets may open more than one gap for low disc viscosities \citep{DongEtal18-alpha0}.

Based on observations of planets at separations less than a few AU, gas giant planets were suggested to not grow via the Core Accretion scenario beyond the gap-opening mass \citep[e.g.,][]{IdaLin04a,IdaLin04b,SantosEtal17,SuzukiEtal18}. Our results extend these suggestions in several ways. Firstly, ALMA planet candidates are separated by $\sim 10-100$~AU from their host stars. Further, their gaseous discs are not yet dispersed, so yield "live" constraints on the process of planet accretion. Finally, most of the planet candidates are not massive enough to open significant gaps in the gas \citep{DipierroEtal15,ClarkeEtal18,DipierroLaibe17}, so their low gas accretion rates cannot be explained by gap opening.

One reason for the inefficiency of gas accretion may be dust opacity. To obtain the envelope contraction time scales within a few Myr, dust opacity is typically assumed to be $\sim 10-100$ times lower than the interstellar dust opacity due to dust growth \citep{PollackEtal96,PN05,Lissauer09,AyliffeBate12,Piso15}. 
However, dust growth could be counteracted by grain fragmentation \citep{DD05,HB11,Mordasini13}, an effect that increases the opacity. Further, composition of gas envelopes of giant planets may be strongly over-abundant in metals and dust for two reasons: (i) the bulk composition of planets is significantly over-abundant in metals compared to their parent stars \citep{MillerFortney11};  and (ii) gas giant planets tend to be found preferentially around metal-rich stars \citep{FischerValenti05}. These two effects may increase envelope dust opacity by a factor of a few to ten compared with that of the interstellar medium at Solar metallicity. \cite{AyliffeBate09b} finds that gas accretion rate onto a $10-50 \mearth$ core is at least an order of magnitude lower for the full insterstellar dust opacity case compared with that for opacity reduced by a factor of 100.

Additionally, classic hydrostatic calculations of envelope contraction \citep[e.g.,][]{IkomaEtal00} assumed 1D geometry. Modern 3D simulations \citep{Szulagyi14,OrmelEtal15,FungChiang16,LambrechtsLega17} show that there is not only an inflow but also an  outflow from the Hill sphere of the planet. This outflow is part of the meridional circulation  between the circumstellar and circumplanetary discs in the case of giant planets \citep{Szulagyi14,FungChiang16}. In the terrestrial regime, a similar recycling of gas within the Bondi-radius has been found \citep{OrmelEtal15}. In both planetary mass regimes, the gas flow enters the Hill sphere from the vertical directions and leaves through the midplane region, leading to a reduction in the gas accretion rate.


\section*{Acknowledgements}
SN acknowledges support from STFC grants ST/N000757/1 and ST/M006948/1 to the University of Leicester. G.D. acknowledges financial support from the European Research Council (ERC) under the European Union's Horizon 2020 research and innovation programme (grant agreement No 681601).
J.Sz. acknowledges the funding from the the Swiss National Science Foundation (SNSF) Ambizione grant PZ00P2\_174115.



\bibliographystyle{mnras}
\bibliography{nayakshin}




\appendix

\section{Data table}

\begin{table}
\centering
\begin{tabular}{lccc}
\hline
Source & Age [Myr] & Sep [AU] & Planet mass [$\mj$] \\
\hline
RY TAU & 1.00 & 43.41 & 0.054 \\
UZ TAUE & 1.26 & 69.00 & 0.009 \\
DS TAU & 3.98 & 32.93 & 6.058 \\
FT TAU & 1.58 & 24.78 & 0.048 \\
MWC 480 & 6.31 & 73.43 & 2.300 \\
DN TAU & 2.51 & 49.29 & 0.005 \\
GO TAU & 5.62 & 58.91 & 0.399 \\
GO TAU & 5.62 & 86.99 & 0.043 \\
IQ TAU & 1.12 & 41.15 & 0.045 \\
DL TAU & 1.00 & 39.29 & 0.091 \\
DL TAU & 1.00 & 66.95 & 0.163 \\
DL TAU & 1.00 & 88.90 & 0.459 \\
CI TAU & 1.58 & 13.92 & 0.750 \\
CI TAU & 1.58 & 48.36 & 0.150 \\
CI TAU & 1.58 & 118.99 & 0.400 \\
HL TAU & 1.41 & 13.20 & 0.200 \\
HL TAU & 1.41 & 32.30 & 0.270 \\
HL TAU & 1.41 & 68.80 & 0.550 \\
AS 209 & 1.10 & 9.00 & 2.050 \\
AS 209 & 1.10 & 99.00 & 0.650 \\
EL 24 & 0.20 & 57.00 & 0.840 \\
EL 27 & 0.79 & 69.00 & 0.060 \\
GW Lup & 2.00 & 74.00 & 0.030 \\
HD 142666 & 12.59 & 16.00 & 0.300 \\
HD 143006 & 3.98 & 22.00 & 20.000 \\
HD 143006 & 3.98 & 51.00 & 0.330 \\
HD 163296 & 12.59 & 10.00 & 0.710 \\
HD 163296 & 12.59 & 48.00 & 2.180 \\
HD 163296 & 12.59 & 86.00 & 0.140 \\
SR 4 & 0.79 & 11.00 & 2.160 \\
GY 91 & 0.25 & 10.40 & 0.113 \\
GY 91 & 0.25 & 40.30 & 0.232 \\
GY 91 & 0.25 & 68.90 & 0.002 \\
TW Hya & 8.13 & 24.00 & 0.013 \\
TW Hya & 8.13 & 41.00 & 0.013 \\
TW Hya & 8.13 & 94.00 & 0.300 \\
HD 169142 & 5.01 & 14.00 & 3.500 \\
HD 169142 & 5.01 & 53.00 & 0.700 \\
PDS 70 & 5.37 & 22.00 & 15.000 \\
\hline
\end{tabular}
\label{table1}
\caption{The data used in this paper. The columns indicate, respectively: (1) star name; (2) age ; (3) gap location; (4) inferred planet mass. See Section~\ref{sec:data} for references and error determination.}
\end{table}



\bsp	
\label{lastpage}
\end{document}